\documentclass[useAMS, usenatbib]{mn2e}

\usepackage{graphicx}
\usepackage{mathptmx}

\newcommand{\civ}{\mbox{C~\textsc{iv}}}

\newcommand{\siiv}{\mbox{Si~\textsc{iv}}}
\newcommand{\nv}{\mbox{N~\textsc{v}}}
\newcommand{\oiv}{\mbox{O~\textsc{iv}]}}
\newcommand{\lya}{Ly$\alpha$}
\newcommand{\kms}{\mbox{km~s$^{-1}$}}
\newcommand{\lam}{$\lambda$}
\newcommand{\aj}{AJ} 
\newcommand{\mnras}{MNRAS} 
\newcommand{\apj}{ApJ} 
\newcommand{\apjl}{ApJ} 
\newcommand{\apjs}{ApJS} 
\newcommand{\aap}{A\&A} 
\newcommand{\araa}{ARA\&A} 
\newcommand{\nat} {Nature} 

\title[Metallicity and Far-Infrared Luminosity of High Redshift Quasars]{Metallicity and Far-Infrared Luminosity of High Redshift Quasars}

\author[L. E. Simon and F. Hamann]{Leah E. Simon$^{1}$\thanks{email: {\tt lsimon@astro.ufl.edu} (LES); {\tt hamann@astro.ufl.edu} (FH)} and Fred Hamann$^{1}$\\
$^{1}$Department of Astronomy, University of Florida, 211 Bryant Space Science Center, Gainesville, FL 32611, USA }

\begin{document}

\pagerange{\pageref{firstpage}--\pageref{lastpage}} \pubyear{2009}

\maketitle
\label{firstpage}

\begin{abstract}
We present the results of an exploratory study of broad line region (BLR) metallicity in 34 2.2~$\le$~z~$\le$~4.6 quasars with far-infrared (FIR) luminosities (L$_{\mathrm{FIR}}$) from 10$^{13.4}$ to $\leq $10$^{12.1}$  L$_{\odot}$.  Quasar samples sorted by L$_{\mathrm{FIR}}$ might represent an evolutionary sequence if the star formation rates (SFRs) in quasar hosts generally diminish across quasar lifetimes.  We use rest-frame ultraviolet spectra from the Sloan Digital Sky Survey to construct three composite spectra sorted by L$_{\mathrm{FIR}}$, corresponding to average SFRs of 4980, 2130 and $\leq 340$~M$_{\odot}$ yr$^{-1}$ after correcting for a nominal quasar FIR contribution.  The measured \nv~\lam1240/\civ~\lam1550 and \siiv~\lam1397+\oiv~\lam1402/\civ~\lam1550 emission line ratios indicate super-solar BLR metallicities in all three composites, with no evidence for a trend with the star formation rate.  The formal derived metallicities, Z$\sim$~5--9~Z$_\odot$, are similar to those derived for the BLRs of other quasars at similar redshifts and luminosities.  These results suggest that the ongoing star formation in the host is not responsible for the metal enrichment of the BLR gas.  Instead, the BLR gas must have been enriched before the visible quasar phase.  These results for high quasar metallicities, regardless of L$_{\mathrm{FIR}}$, are consistent with evolution scenarios wherein visibly bright quasars appear after the main episode(s) of star formation and metal enrichment in the host galaxies.  Finally, young quasars, those more closely associated with a recent merger or a blowout of gas and dust, may exhibit tracers of these events, such as redder continuum slopes and higher incidence of narrow absorption lines.  With the caveat of small sample sizes, we find no relation between L$_{\mathrm{FIR}}$ and the reddening or the incidence of absorption lines.

\end{abstract}


\section{Introduction}
\label{sec:intro}

Numerous studies have shown a strong correlation between supermassive black hole (SMBH)  mass and host galaxy mass over broad mass ranges, implying that there is a close evolutionary relationship between them \citep{Gebhardt00, Merritt01, Tremaine02, Marconi03, Haring04, Shields06}.  Luminous quasars at high redshifts, which represent the most massive SMBHs, are experiencing vigorous accretion of a significant portion of the final SMBH mass. This accretion activity is believed to be triggered by global processes (galaxy mergers, interactions or perhaps secular evolution) that also trigger major episodes of star formation in the massive host galaxies, which are also rapidly being assembled at high redshifts.  Therefore, the processes of SMBH formation and growth resulting in the quasar phenomenon are directly linked to the birth of massive galaxies \citep{Haehnelt98, Richstone98, Omont01, Omont03,  Beelen06, Cox06}.  However, the nature of the relationship between quasars and galaxy formation is not well understood.

It is widely believed that strong interactions between gas rich galaxies can result in ultra-luminous infrared galaxies (ULIRGs), defined by L$_{\mathrm{IR}}~>$10$^{12}L_{\odot}$, which have star formation rates (SFRs) of $>~100~M_{\odot}~yr^{-1}$ \citep{Houck85, Omont01, Omont03, Flores04, Cox05, Beelen06, Daddi07, Cao08}.  A fraction of ULIRGs have been found to contain dust-enshrouded quasars \citep{Sanders96, Lonsdale06}.  There is evidence at both low and high redshift that these embedded quasars are precursors to optically luminous quasars.  For example, \citet{Cao08} find that low redshift quasars with IR luminosities in the ULIRG range fall between optically selected PG quasars and ULIRGs (starbursts) for a variety of mid-IR spectroscopic indicators of starburst and active galactic nucleus (AGN) contributions including: polycyclic aromatic hydrocarbon (PAH) luminosities, fine structure emission line strengths, silicate absorption strengths, spectral slope and mid-IR colour indices.  These findings suggest that IR luminous quasars could represent an intermediate stage between a dominant starburst and a dominant AGN phase.  

Quasars with high SFRs therefore may be at an intermediate stage between an embedded quasar in a star forming galaxy and a visible quasar in a galaxy where most star formation has ceased (see also \citet{Sanders88}).  At low redshifts, numerous star formation indicators in quasar host galaxies are seen, e.g. high IR luminosities, PAH emission, strong (sub)-mm emission and strong CO emission \citep{Hao05, Schweitzer06, Farrah07a, Farrah07b, Netzer07}.   At high redshifts (z~$\sim$~2-6), observations at sub-mm and mm wavelengths (rest-frame far-infrared (FIR) to sub-mm, depending on wavelength and redshift range) of optically luminous quasars suggest that up to 30\% of quasars also fall within the ULIRG range, similar to the IR quasars studied by \citet{Cao08} \citep{Carilli01, Omont01, Omont03, Cox05, Beelen06, Hao08}.  \citet{Coppin08} compare dynamical, gas and SMBH masses of ten z~$\approx 2$~ sub-mm detected quasars and z~$\approx 2$~ sub-mm galaxies (SMGs), which are high redshift counterparts to ULIRGs, though less extreme with more evenly distributed star formation instead of a localised starburst \citep{Menendez-Delmestre09}. They find that the fainter half of their quasar sample could likely be `transition objects' between SMGs and luminous quasars, based on their SMG-like surface densities and their proximity to the local M$_{\mathrm{BH}}$/M$_{\mathrm{sph}}$ relation, since luminous quasars tend to lie above this relation and typical SMGs tend to lie below.  The highest redshift sub-mm-selected source currently known has also been observed to posses similar stellar and gas masses to this z~$\approx 2$~ sample of transition objects, as well as a small AGN contribution \citep{Coppin09}.  These quasars eventually may produce enough energy from accretion to effectively blow out the surrounding gas, halting star formation and clearing the view to an optically bright quasar before halting their own growth \citep{Wyithe03, DiMatteo05, Hopkins08}.  This regulation of star formation by the accreting SMBH could naturally produce the observed black hole-galaxy mass correlation, and different stages of the process should present different SFRs, with initially high SFRs declining as the AGN becomes more dominant, culminating in a phase of strong SMBH accretion and little or no ongoing star formation \citep{Kauffmann00, Granato04}.  

FIR luminosities can be used to determine the star formation rates in quasar host galaxies.  The FIR emission is due to dust heated either by star formation or quasar emission (see \citet{Haas03} for discussion).  \citet{ Lutz07, Lutz08} find evidence that FIR emission in quasar host galaxies at all redshifts is caused by dust heated by star formation and not the quasar.  The strength of PAH emission, which is found almost exclusively in star forming regions, tightly corresponds to the strength of FIR emission in the same hosts \citep{Dale01, Calzetti07, Lutz08}.  \citet{Beelen06} measure FIR emission from six high redshift quasars and derive FIR to radio spectral indexes consistent with local star forming galaxies without AGN. The FIR emission is seemingly uncontaminated by hotter dust potentially heated by the AGN.  These results along with others (see for example,  \citet{Efstathiou95, Serjeant09}) further substantiate the claim that the FIR luminosity in quasars is dominated by star formation and not by the AGN.

FIR luminosity traces ongoing star formation, but past star formation also can be observed indirectly by measuring chemical abundances.  Several studies have found that high-redshift quasars typically have metallicities greater than or equal to solar metallicity in the broad emission line region (BLR), which requires significant previous star formation in the host \citep{Hamann99, Dietrich03, Warner04, Nagao06a}. These BLR studies rely on emission line ratios such as \nv/\civ~ and \siiv+\oiv/\civ~ that are sensitive to the selective enrichment of nitrogen as a secondary element and/or to the decreasing temperatures and increasing metal-line saturations that occur in metal-rich gas.  The main result for typically solar or higher metallicities near quasars has been corroborated by independent studies of the narrow emission lines \citep{Groves06, Nagao06b} and narrow absorption lines \citep{Hamann99, Dodorico04, Gabel06, Simon10} in quasar spectra.  The metal-rich BLR result is true even for the highest redshifts studied, e.g. \citet{Pentericci02, Jiang07, Juarez09}, with redshifts out to z~=~6.4. There is no known change in metallicity with redshift (see \citet{Matsuoka09} and references above).  Simple chemical evolution models for quasars and elliptical galaxies find that galactic centres tend to be more metal-rich than their halos, and a centralised starburst can enrich the galactic centre to supersolar abundances in a short time ($\le 10^8~yr$) \citep{Friaca98, Hamann99, Granato01, Hamann02, Granato04, Hamann07, Juarez09}.  These models, combined with the consistently super-solar gas abundances observed in quasar environments, imply that quasars tend to emerge after or near the end of (potentially) short centralised star formation epochs.  If the quasar phase emerges when star formation is on the decline, less advanced environments might have higher SFRs and lower abundances \citep{Georgakakis09}.

The absorption lines in quasar spectra provide additional information about outflows that might be related to the blowout of gas from the host galaxies, and perhaps, about the gaseous remnants of recent galaxy mergers.  These absorption lines may be more common in quasars with higher SFRs if mergers and interactions trigger the star formation as in ULIRGs \citep{ Weymann91, Becker00, Richards01a, Rupke05, Georgakakis09}.  Quasar `associated' \civ~ absorption lines (AALs), near the emission redshifts with velocity widths less than 500~\kms and broad absorption lines (BALs), with velocity widths greater than a few thousand \kms, are examples of these potential merger, interaction and outflow signatures.  In systems with recent mergers, remnants from the interaction may manifest as a greater incidence of low-velocity ($< 2000$~\kms) AALs.  About 25\% of bright quasars contain AALs with rest equivalent widths (REWs) of $>$0.3~\AA, and around 10\% of quasars have BALs \citep{Ganguly01, Vestergaard03, Misawa03, Trump06, Nestor08, Wild08, Gibson08, Paola08}.  A higher incidence of AALs or BALs may occur in the quasars with higher SFRs if these quasars have more recently experienced an interaction and/or there is a progression in quasar outflow characteristics with time. 

In this paper, we present an exploratory observational study examining whether SFR in the host galaxies correlates with metallicity in the near-quasar environment.  We measure metallicity in the quasar BLR from the rest frame ultraviolet (UV) spectrum and estimate galactic SFRs from the FIR luminosities.

The data and analysis are described in \S~\ref{sec:data} and \S~\ref{sec:anal}.  The results and discussion are presented in \S~\ref{sec:disc}, with a brief summary in \S~\ref{sec:sum}.  We adopt cosmological parameters H$_0~=~70$~\kms~Mpc$^{-1}$, $\Omega_m~=~0.3$, and ~$\Omega_{\Lambda}~=~0.7$ throughout this work.  

\section{Data}
\label{sec:data}

We select quasars for this study from the sample observed with MAMBO at IRAM at 1.2~mm by \citet{Carilli01} and \citet{Omont01, Omont03} and SCUBA at JCMT at 850$\mu m$ by \citet{Isaak02}, \citet{McMahon99} and \citet{Priddey03} and all compiled by \citet{Hao08}.  The quasars all were selected to be optically bright with absolute B-band Magnitude M$_B < -26.1$ for the Carilli et al. objects, M$_B < -27.0$ for the Omont et al. objects and M$_B < -27.5$\footnote{They use an Einstein deSitter cosmology, $\Omega_M = 1, \Omega_\Lambda = 0, H_0 = 50$~\kms.} for the \citet{McMahon99, Isaak02} and \citet{Priddey03} objects.  \citet{Carilli01} observed a representative sample of 41 out of the more than 100 quasars with z~$\ge$~3.6 found as part of the Sloan Digital Sky Survey (SDSS) Galactic Cap and Southern Equatorial Stripe survey.  \citet{Omont01, Omont03} observed a random selection of 97 radio quiet, optically luminous sources from the multicolour Palomar Digital Sky Survey available from G. Djorgovski's web page\footnote{http://astro.caltech.edu/\~{}george/z4.qsos} with 3.9~$<$~z~$<$~4.5 and the \citet{Veron-Cetty00} catalogue with 1.8~$<$~z~$<$~2.8.   \citet{McMahon99} selected a small sample of 6 bright $z > 4$ radio quiet quasars from the APM survey \citep{StorrieLombardi96}.  \citet{Isaak02} selected a larger sample of the 76 most UV luminous $z \ge 4$ radio quiet quasars known at the time of observation, and \citet{Priddey03} selected a complimentary sample of 57 $z \ge 2$ quasars from various large surveys. 

We cross-reference this sample with the optical quasar spectra in the SDSS data-release 6, finding 116 objects with available spectra.  The SDSS spectra from data release 6 have resolution R~$=~\lambda/\delta\lambda~\sim$2000 and wavelength coverage $\lambda~=3800-9200~\AA$ \citep{Adelman08}.  We consider only those objects with redshifts between 2.17 and 4.75, compatible with SDSS spectral coverage of the full 1200 -- 1600~\AA~ rest-frame wavelength range. Spectra missing regions within the specified wavelength range are further excluded from the analysis.  Five spectra with very low signal to noise ratios (S/N) also are rejected. The final sample consists of 34 optical SDSS spectra with a range of sub-mm brightnesses and a redshift range of 2.2~$<~z~<$~4.6.

\begin{table*}
\centering
\begin{minipage}{11cm}
\centering
\caption{Quasar Sample}
\begin{tabular}{llcccccc}
\hline
\label{tab:qso}
& & z & $\log \mathrm{L_{bol}}$ & M$_B$ & $\log \mathrm{L_{60\mu m}}$ &  SFR &  \\
& Name &  & log(erg s$^{-1}$) & & $\log \mathrm{L}_{\odot}$ & M$_\odot \mathrm{yr}^{-1}$ & Ref.$^a$\\ 
\hline
& FIR Bright & & & & & & \\
& PSS J1057+4555 & 4.12 & 48.1 & -29.0 & 13.2 & 3720 & O01\\
& PSS J1347+4956 &  4.56 & 47.9 & -28.5 & 13.3 & 4690 & O01\\
& J144758.46-005055.4  & 3.80 & 47.2 & -26.8 & 13.3 & 6350 & C01\\
& J154359.3+535903 & 2.37 & 47.9 & -28.4 & 13.3 & 5550 &O03\\
& PSS J1248+3110  & 4.35 & 47.6 & -27.8 & 13.3 & 6220 & O01\\
& PSS J1418+4449  & 4.28 & 48.0 & -28.8 & 13.3 & 5320 & O01\\
& PSS J0808+5215 & 4.45 & 48.1 &-28.9 & 13.3 &  5350 & O01\\
& J110610.8+640008 & 2.19 & 48.3 & -29.4 & 13.4 & 4720 &O03\\
& HS B1140+2711  & 2.63 & 48.0 & -28.6 & 13.2 &  3290 & P03\\
& J142647.82+002740.4 & 3.69 & 47.3 & -26.8 & 13.2 & 4580 & C01\\
\hline
& FIR Intermediate & & & & & &\\
& J023231.40-000010.7 & 3.81 & 47.2 & -26.7 & 12.8 & 1870 &C01\\
& PSS J1535+2943 & 3.99 & 47.4 & -27.3 & 12.8 & 1710& O01\\
& J012403.78+004432.7  & 3.81 & 47.9 & -28.4 & 12.9 & 1230 &C01\\
& J025518.58+004847.6 & 3.97 & 47.6 & -27.8 &12.9 & 1670 &C01\\
& J015048.83+004126.2  & 3.67 & 47.7 & -27.9 & 12.9 & 1920 &C01\\
& J141332.35-004909.7  & 4.14 & 47.4 & -27.3 & 12.9 & 2330 &C01\\
& J025112.44-005208.2  & 3.78 & 47.3 & -26.9 & 13.0 & 2580 & C01\\
& J111246.30+004957.5  & 3.92 & 47.6 & -27.7 & 13.0 & 2530 & C01\\
& PSS J1317+3531 & 4.36 & 47.6 & -27.7 & 13.1 & 3330 & O01\\
\hline 
& FIR Faint & & & & &&\\
& J015339.61-000910.3  & 4.20 & 47.5 & -27.4 & $\leq$12.5 &  $\leq$430 &C01\\
& J140554.07-000037.0  & 3.55 & 47.5 & -27.6 & $\leq$12.7 &  $\leq$850 &C01\\
& J225419.23-000155.0  & 3.68 & 47.3 & -26.9 & $\leq$12.1 &  $\sim$0 &C01\\
& PSS J1443+5856  & 4.27 & 48.0 & -28.7 & $\leq$12.4 &  $\sim$0 &O01\\
& J225759.67+001645.7  & 3.75 & 47.5 & -27.4 & $\leq$12.2 & $\sim$0 & C01\\
& HS B0808+1218  & 2.26 & 47.6 & -27.8 & $\leq$12.4 &  $\sim$0 & P03\\
& HS B1111+4033  & 2.18 & 47.7 &-27.9 & $\leq$12.7 &  $\leq$550 & P03\\
& LBQS B1210+1731  & 2.54 & 47.8 &-28.1 & $\leq$12.4 &  $\sim$0 & P03\\
& LBQS B1334-0033  & 2.80 & 47.6 &-27.8 & $\leq$12.6 &  $\leq$610 & P03\\
& BR 1600+0729  & 4.35 & 47.8 &-28.3 & $\leq$12.7 &  $\leq$280 & M99\\
& PSS J0852+5045  & 4.2 & 47.7 &-28.0 & $\leq$12.4 &  $\sim$0 & I02 \\
& PSS J0957+3308  & 4.25 & 47.8 &-28.1 & $\leq$12.7 &  $\leq$650 & I02 \\
& PSSJ1618+4125  & 4.21 & 47.4 & -27.2 & $\leq$12.7 &  $\leq$990 & O01\\
&PSS  J1633+1411  & 4.35 & 47.6 & -27.6 & $\leq$12.6 &  $\leq$570 & O01\\
& J142329.98+004138.4  & 3.76 & 47.2 & -26.6 & $\leq$12.2 &  $\leq$180 & C01\\
\hline
\end{tabular}
\footnotetext[0]{$^a$O01 \citet{Omont01}, O03 \citet{Omont03}, I02 \citet{Isaak02}, C01 \citet{Carilli01}, P03 \citet{Priddey03}, M99 \citet{McMahon99}}
\end{minipage}
\end{table*}

The MAMBO 1.2~mm and SCUBA 850~$\mu$m observations correspond to 160 -- 400~$\mu$m rest wavelengths, depending on quasar redshift.  We follow the prescription in \citet{Hao08} to convert from observed flux into FIR luminosity at 60~$\mu m$, L$_{60}~=~\lambda L_{\lambda}(60~\mu m)$, in which we assume that a grey-body spectrum describes the rest frame FIR spectral energy distribution (SED) with a dust temperature of 41~K and dust emissivity index of 1.95, as detailed in \citet{Priddey01}. We find agreement with \citet{Hao08} to within 10\%.  

\begin{figure}
\centering
\vspace{4mm}
\includegraphics[width=8cm]{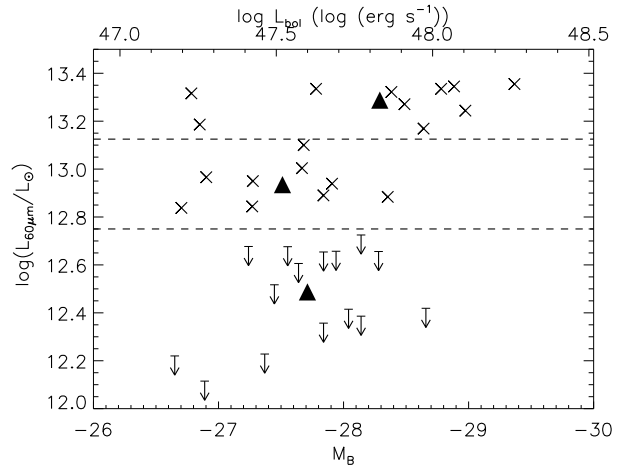}
\caption{Quasar FIR luminosities, absolute B magnitude and bolometric luminosities.  Horizontal dashed lines represent the division between the three FIR luminosity bins, crosses are FIR-detected quasars and arrows are upper limits. Filled triangles are average luminosities for each FIR luminosity bin.}
\label{fig:l_m}
\end{figure}

The L$_{60}$, absolute B-band magnitude (M$_B$) and bolometric luminosity (L$_{\mathrm{bol}}$) for each object are shown in Figure~\ref{fig:l_m}, and listed in Table~\ref{tab:qso}.  The FIR luminosity, log(L$_{60}/\mathrm{L}_{\odot})$, ranges from $\leq$12.1 to 13.4, is dominated by star formation and falls roughly within the ULIRG range.  We estimate the star formation rate of each quasar host from L$_{60}$, corrected to exclude the small quasar contribution by assuming the hosts follow the same regression line as typical quasars for L$_{\mathrm{bol}}$ vs. L$_{60}$ as shown in Figure 1 of \citet{ Hao08}.  This corrected L$_{60}$ is then used in Hao et al.'s equation 2, which is derived from the Kennicutt star formation rate law \citep{Kennicutt98}, and the estimated star formation rates for each object are also listed in Table~\ref{tab:qso} .   We convert the M$_B$ from the observation papers to our adopted cosmology, and obtain M$_B$ spanning the range of -26.7 to -29.4, which corresponds to log(L$_{\mathrm{bol}}$) of 47.2 to 48.3~log(erg~s$^{-1}$).  L$_{\mathrm{bol}}$ measures the quasar black hole accretion luminosity, and is estimated using a bolometric correction factor of 9.74 applied to the monochromatic continuum luminosity \lam L$_{\lambda}$(4400~\AA) ($\sim M_B$) following \citet{Vestergaard04}. 

We separate the sample into three bins based on L$_{60}$ such that each bin contains a similar number of objects: FIR bright quasars with log(L$_{60}$/L$_{\odot}$)~$\geq$~13.17, FIR intermediate quasars with 12.8~$<$~log(L$_{60}$/L$_{\odot}$)~$\leq$~13.1 and FIR faint quasars with log(L$_{60}$/L$_{\odot}$)~$<$~12.75, in which the FIR faint luminosities are upper limits.  Any objects with log(L$_{60}$/L$_{\odot}$) upper limits above 12.75 were not included in the study.  The FIR-bright and FIR-faint bins each span the redshift range from $\sim$2.2 to $\sim$4.4.  The FIR bright bin also includes one object with z~$\sim$4.6, and the FIR intermediate bin spans a smaller redshift range from 3.7 to 4.4.  The divisions between the L$_{60}$ bins are represented in Figure~\ref{fig:l_m} by horizontal dashed lines and average values in each bin for  L$_{\mathrm{bol}}$~ and L$_{60}$~ are denoted by filled triangles.  These average values are listed in Table~\ref{tab:abs} for each luminosity bin in columns 2 and 3, along with the corresponding star formation rates in column 4, and several abundance indicators, discussed in \S's~\ref{subsec:metal} and \ref{sec:disc}.  


\begin{table*}

\centering

\caption{Metallicity from Emission Line Flux Ratios}
\begin{tabular}{llccccccl}
\hline
\hline
\label{tab:abs}
& Composite spectra & $<$log L$_{\mathrm{bol}}>$ & $<$log L$_{60}>$ & $<SFR>$ & Flux Ratio & Z/Z$_{\odot}$  & Flux Ratio & Z/Z$_{\odot}$ \\
& & log(erg~s$^{-1}$)& log(erg~s$^{-1}$)& M$_{\odot} $yr$^{-1}$ & \nv/\civ & \nv/\civ & \siiv+\oiv/\civ & \siiv+\oiv/\civ \\
\hline
& FIR Faint & 47.60 & $\leq$46.07 & $\leq$340 & 1.2 & 7.9 & 0.3 & 3.0 \\
& FIR Int. & 47.52 & 46.52 & 2130 & 1.4 & 9.6 & 0.3 & 3.0 \\
& FIR Bright & 47.89 & 46.87 & 4980 & 1.5 & 10.9 & 0.4 & 6.5 \\
\hline
\end{tabular}

\end{table*}

\section{Analysis}
\label{sec:anal}
\subsection{Composite Spectra}

To make comparisons between different L$_{60}$, we create one composite spectrum from the SDSS spectra comprising each of the three FIR bins in the sample described in \S~\ref{sec:data}.  To create the composites, we begin by shifting each spectrum to its rest wavelength using the redshifts provided by SDSS \citep{Subbarao02}.  We manually inspect each individual spectrum and remove absorption features by interpolating across the affected regions.  The presence of narrow absorption lines and their relationship to L$_{60}$ is discussed in \S~\ref{subsec:aalanal} below.  Then we average together the spectra in each bin to create the final composites.  To ensure that no single spectrum is dominating the composite we also compute the median spectra for each bin, compare them to the average spectra used throughout the rest of the analysis, and confirm that the two composite types are well-matched.  

Our analysis is limited to the spectral region between \lya~1216~\AA~ and \civ~1550~\AA.  These limits are imposed by the incomplete spectral coverage at longer wavelengths and by suppression in the \lya~ forest starting in the blue wing of the \lya~ emission line and extending to shorter wavelengths.  Before normalising the composite spectra, we visually compare the continuum slopes of the three composites.  We find no significant trend for reddening with FIR luminosity in the continuum slopes of the composite spectra.  Differences between the composite slopes are negligible compared to the dispersion of slopes among the individual quasars making up each composite.  

\begin{figure*}
\centering
\vspace{4mm}
\includegraphics[width=16cm]{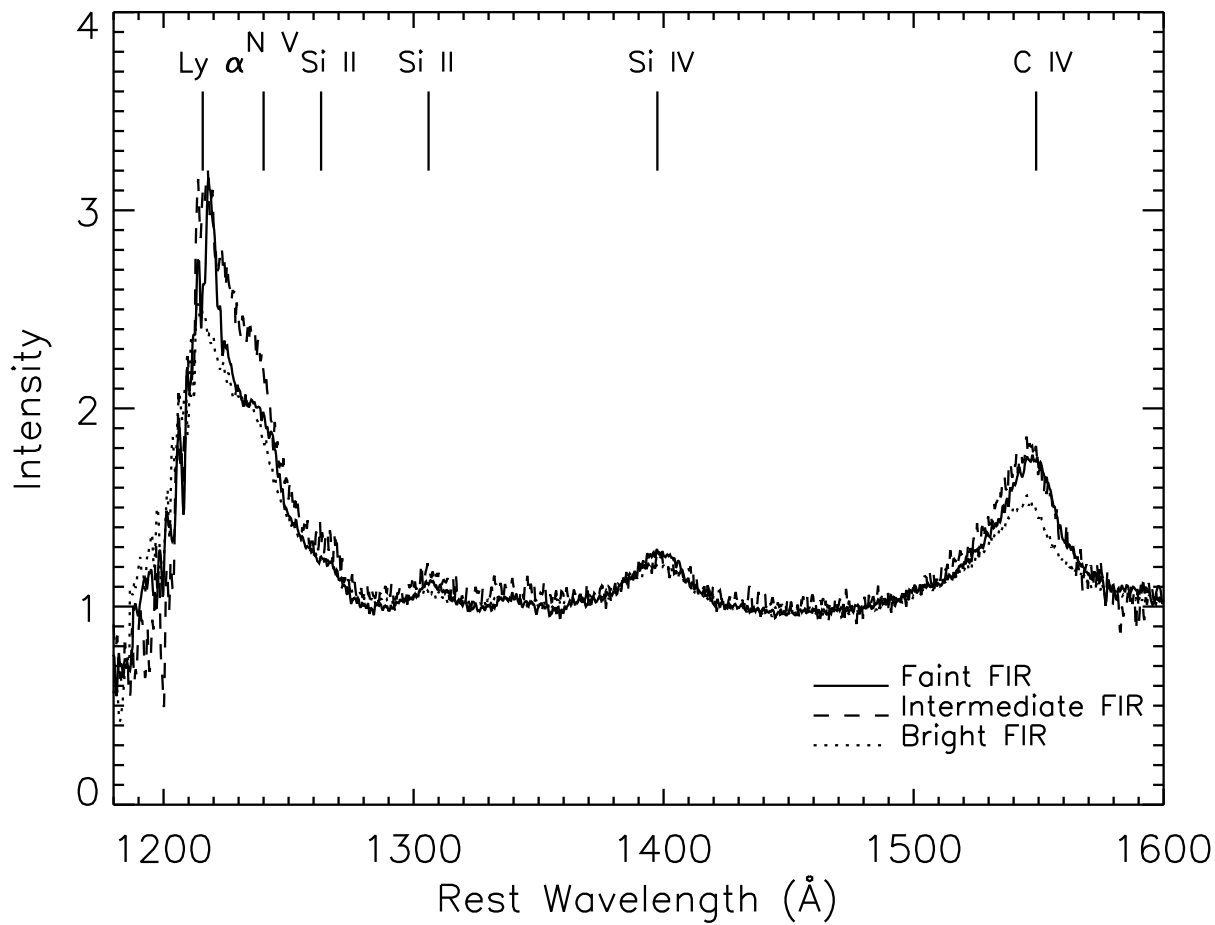}
\caption{Normalised composite spectra for FIR-faint (solid line), FIR-intermediate (dashed line) and FIR-bright (dotted line) for the (rest) wavelength range from \lya~ to \civ.  Prominent emission features are labelled. The continuum has been normalised to 1, as described in the text. }
\label{fig:comp}
\end{figure*}

We fit a power-law continuum to the FIR-intermediate composite spectrum using wavelength regions devoid of absorption or emission, 25~\AA~ wide and centred at ~\lam1460, and ~\lam1770, following \citet{Warner03}.  Unfortunately, the FIR-faint and FIR-bright composites have incomplete spectral coverage past $\sim$\lam1700~\AA, so we fit the continuum using the available wavelength region around ~\lam1460~\AA, plus the region from ~\lam1335-1355~\AA, which is also used by \citet{Juarez09}, and which we determine by visual inspection to be devoid of absorption or any detectable emission.  The normalised average composite spectra are shown in Figure~\ref{fig:comp}.

\subsection{Emission Line Flux Ratios}
\label{subsec:ratioanal}

We measure several quasar broad emission line fluxes to calculate line flux ratios, which we then use to estimate the gas phase metallicity in the near-quasar environment.  Flux ratios are measured for the emission lines that have been shown by e.g. \citet{Hamann02, Juarez09}, to give good abundance estimates within the limited wavelength range from ~\lam1216~\AA~ to ~\lam1600~\AA: \nv~\lam1240/\civ~\lam1549 and \siiv~\lam1397+\oiv~\lam1402/\civ~\lam1549.  These emission lines are labelled in Figure~\ref{fig:comp}.  Other ratios such as N~III]~\lam1750/O~III]~\lam1664, \nv~\lam1240/He~II~\lam1640 or N~III~\lam991/C~III~\lam977 are too weak given the S/N in the spectrum and/or they fall in a poorly characterised region of the spectrum.  The measured line flux ratios are listed in columns 5 and 7 of Table~\ref{tab:abs}.  
 
\begin{figure}
\centering
\vspace{4mm}
\includegraphics[]{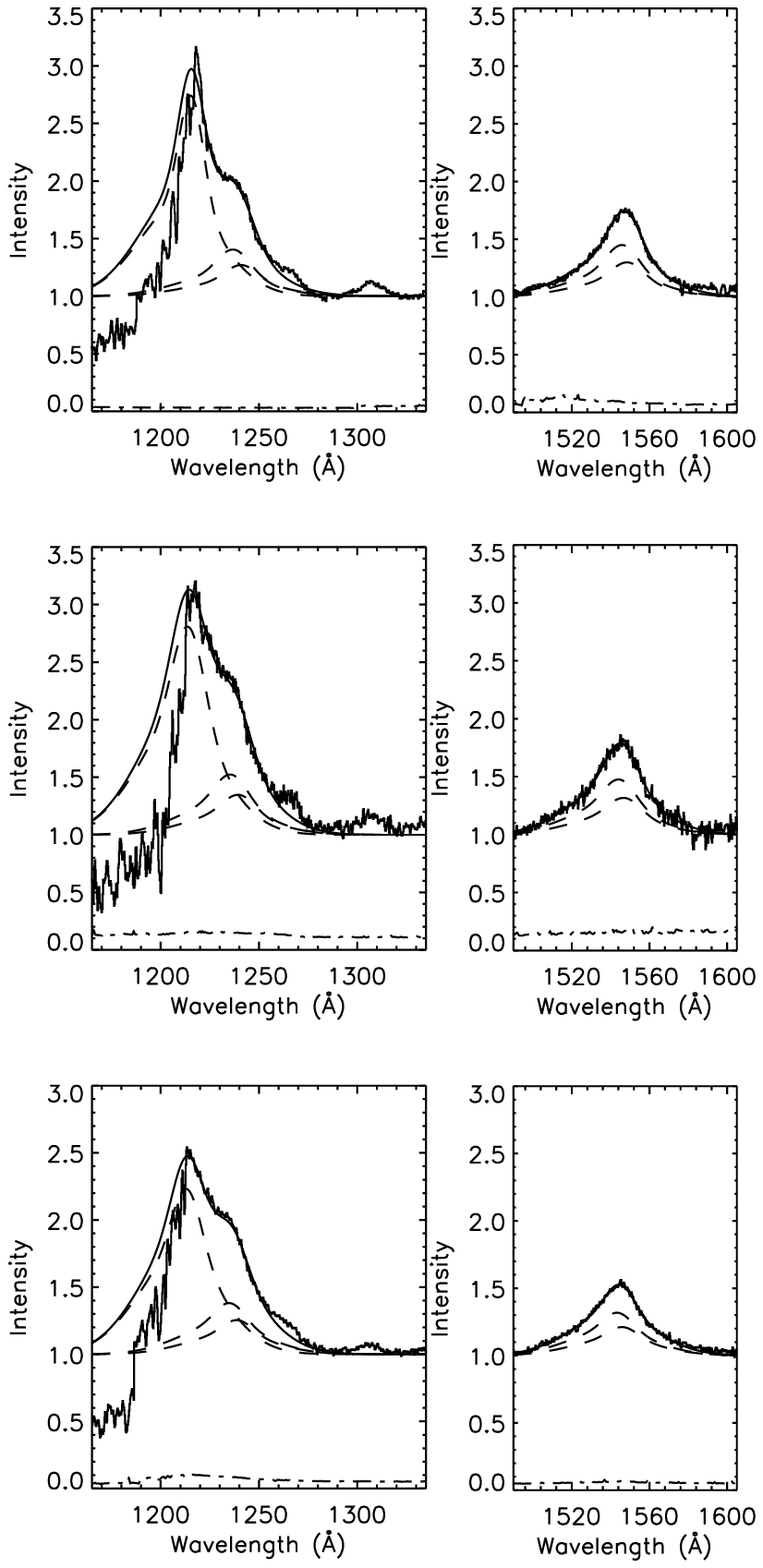}
\caption{Gaussian fits for \lya, and \nv~ in the left panels and \civ~ in the right panels for each normalised composite spectrum. The vertical scale is normalised flux.  The top panels shows the FIR faint spectrum, the middle panels shows the FIR intermediate spectrum and the FIR bright spectrum is shown in the bottom panels.  The solid curves are the data and total gaussian fits, the dashed curves show the individual gaussian fits for each emission line.  The \nv~ and \civ~ features each have two gaussians, one for each doublet member, which are each composed of a narrow and a wide gaussian (not shown). The error spectra for each region are shown as dot--dashed lines. }
\label{fig:gauss}
\end{figure}

Because the \nv~ emission is strongly blended with the \lya~ emission, which is itself significantly degraded by absorption in the \lya~ forest, \civ~ is the only strong, non-blended emission line in the spectra.  Thus, the \civ~ emission lines are fit with gaussians and used as templates for the other emission lines.  The fits to \civ~ and the scaled fits to \lya~ and \nv~ are shown for the three composites in Figure~\ref{fig:gauss}.  The fits are performed using \textsc{GATORPLOT}, an \textsc{IDL} program written by C. Warner\footnote{http://www.astro.ufl.edu/\~{}warner/GatorPlot/}, and are carried out using the smallest number of gaussians possible to provide a good match to the data: one broad and one narrow component for each of the two doublet lines at \lam1548 and \lam1551~\AA.  The two gaussians comprising the \lam 1548 emission line fit are held at fixed full width at half maximum (FWHM) and are scaled in total flux at the fixed central wavelength of the \lya~\lam1216~\AA~ and \nv~\lam1239, 1243~\AA~ doublet emission to match these line strengths.   Because \civ~ is known to be blue-shifted relative to \lya~ and the nominal quasar redshift by $\approx$~310~km~s$^{-1}$ \citep{Tytler92}, the locations of the \nv~ and \lya~ centroids are shifted from their laboratory values by this amount relative to \civ.  

To accurately determine the \nv~ flux, the red \lya~ wing beneath the \nv~ emission must be characterised.  To do this precisely, we scale the \civ~ fit to align with the non-absorbed region of \lya, which in some cases is quite small ($\le 20$~\AA).  The scaled fit may even rise above the data where heavy \lya~ forest absorption has eaten into the emission, as is the case for all of the three composites, or miss the peak of \lya~ as is the case in the FIR faint composite; both cases are clearly shown in Figure~\ref{fig:gauss}.  In order to better match the \lya~ emission peak in the FIR faint composite, we allow the \lya~ gaussian to be $\approx 15\%$ narrower than the \civ~ gaussian.  Narrowing the gaussian for the \lya~ emission line does not have a noticeable effect on the resulting \nv~ strength.  Regardless, a precise match to the \lya~ emission line is not required for this analysis, and does not effect the final \nv~ emission results (see e.g. \citet{Baldwin03} for further discussion).

The \civ~ and \nv~ doublets are fit at their respective fixed separations and with a 3:2 intensity ratio (halfway between the allowed 2:1 and 1:1 intensity ratios) for the shorter and longer wavelength respectively, as in \citet{Baldwin03}.   Possible broadening in \nv~ is predicted if the \nv~ emission forms at a larger distance from the quasar than the \civ~ emission.  \citet{Peterson04} and \citet{Peterson08} find that \nv~ forms twice as close to the quasar as \civ, so if the line widths are controlled by virial motions, \nv~ could be up to $\sim\sqrt{2}$ times broader than \civ.  \lya, on the other hand, forms at about the same distance from the quasar as \civ~ and should be no broader (and probably narrower) than \civ~ \citep{Peterson04, Peterson08}.  Therefore, the \nv~ profiles we adopt, which have the same FWHM's as \civ, may be narrower than the actual \nv~ FWHM's, while the \lya~ FWHM could be slightly broader.  We estimate the uncertainties in the \nv~ line strengths due to uncertainty in the continuum placement by repeating the line fits with different reasonable continuum heights.  We also include estimates of \nv~ line strength uncertainties due to blending with \lya~ and the poorly defined \nv~ FWHM by repeating the line fits with a range of different \nv~ and \lya~ FWHM's.  The overall uncertainties, combining these effects, are less than a factor of 1.5, so that the measurement errors in the \nv/\civ~ ratios are also less than a factor of 1.5.

The \siiv+\oiv~ emission line is weaker than the \civ~ and \nv~ emission lines in the composite spectra, however it also is isolated from other emission.  This isolation means that measuring the equivalent width of the emission without fitting, using \textsc{IRAF}\footnote{\textsc{IRAF} is the Image Reduction and Analysis Facility, supported by NOAO and AURA Inc.}, produces accurate fluxes.  We confirm the accuracy of the equivalent width measurements by comparing the previously measured fluxes from gaussian fits to \civ~ to the fluxes from \civ~ equivalent widths, and find compatible results.  We measure \siiv+\oiv~ and \civ~ equivalent widths to calculate the \siiv+\oiv/\civ~ line flux ratios listed in Table~\ref{tab:abs}. To estimate uncertainties in the \siiv+\oiv~ line strengths, we vary the equivalent width measurements by using a range of continuum heights and wavelength cutoffs for the edge of the emission lines.  The overall uncertainties in the \siiv+\oiv~ line strengths are less than a factor of 1.2, so that the measurement errors in the \siiv+\oiv/\civ~ ratios are also less than a factor of 1.2.   

\subsection{Metallicity}
\label{subsec:metal}

We convert the \nv/\civ~ flux ratios into metallicities using the theoretical relationship in which nitrogen abundance increases relative to carbon as metallicity increases because of secondary enrichment (CNO nucleosynthesis) processes, and determine the average metallicity for each L$_{60}$ composite \citep{Shields76, Hamann93b, Hamann99}.  The correlation is characterised by \citet{Hamann02} with recent updates to the solar abundance ratios by \citet{Dhanda07}.  We find supersolar metallicities for all three composite spectra, as listed in column 6 of Table~\ref{tab:abs}.  

Similar characterisations for the \siiv+\oiv/\civ~ flux ratio-metallicity correlation are performed by \citet{Nagao06a}, using the older solar abundances also used by \citet{Hamann02}.  We apply the latest corrections for the solar abundance ratios for this characterisation and determine that the metallicities are also supersolar according to this ratio, as shown in column 8 of Table~\ref{tab:abs}.  

There is no significant trend among the three composites in the line ratios, based both on our measurements of these ratios and on a visual inspection of the stacked composites in Figure~\ref{fig:comp}, and correspondingly there is no significant trend in metallicity.  The average metallicities across the three composites are Z$\sim$9.5~Z$_{\odot}$ for the \nv/\civ~ ratio and 4.2~Z$_{\odot}$ for the \siiv+\oiv/\civ~ ratio.  

We note that the metallicities inferred from the two line ratios can differ by as much as a factor of $\sim$2 in the same spectrum.  This is typical of the dispersion found between different line ratios in other studies, and it might be a good indication of the theoretical uncertainties \citep{Dietrich03, Nagao06a, Hamann09}.  We compare our metallicities to the metallicities in the L$_{\mathrm{bol}}$ range 10$^{47}$ to 10$^{48}$~erg s$^{-1}$ sampled by a large emission line study by \citet{Warner04}, who find metallicities of Z~$\sim$3-6~Z$_{\odot}$, broadly consistent with the metallicities found in this sample.  We also note that the results based on \nv/\civ (and \nv/He~\textsc{ii}) tend to be higher than other line ratios, and thus our best guess at the metallicity in the current sample overall would be near the lower end of the range 4--9 Z$_{\odot}$ \citep{Hamann02, Baldwin03, Nagao06a}.  The \emph{relative} metallicities between the three composites, which smooth over object-to-object scatter, are robust for differences greater than a factor of 2-3, based on the uncertainties in the line ratio measurements plus the theoretical uncertainties, and therefore useful for spotting strong metallicity trends among the different L$_{60}$ bins.

\subsection{Absorption Lines}
\label{subsec:aalanal}

We inspect each SDSS spectrum used to make the composites and count the number of \civ~ AALs with REW~$> 0.3$~\AA~ per spectrum.  We find that 17/34 quasars have at least one AAL and 6/34 have absorption too broad to be classified as an AAL, yet too narrow to be classified as a BAL, with widths more than twice the thermal width and velocity dispersion expected for unrelated gas in the line of sight, which is indicative of gas forming in an outflow \citep{Bahcall67, Young82}.  These statistics are broadly consistent with previous work, given the small numbers involved in this study \citep{Vestergaard03, Trump06, Nestor08, Wild08, Gibson08, Paola08}.   AALs and outflow lines appear in each of the three L$_{60}$ bins, with no significant differences in their occurrence fractions among the bins that would indicate a trend with L$_{60}$, particularly given the small number of objects in each bin. 
 
\section{Discussion}
\label{sec:disc}

The average line flux ratios \nv/\civ~$\approx~1.35$  and \siiv+\oiv/\civ~$\approx 0.33$~ across the three composites, as shown in columns 5 and 7 of Table~\ref{tab:abs}, correspond to average metallicities of Z~$\sim$~9.5 and 4.2~Z$_{\odot}$ respectively, shown in columns 6 and 8 of Table~\ref{tab:abs}.  There is no significant trend in metallicity with L$_{60}$ in this sample.  We note that the FIR bright bin has the most luminous average L$_{\mathrm{bol}}$, while the FIR faint and intermediate bins have roughly equal, and less luminous average L$_{\mathrm{bol}}$.  This difference in L$_{\mathrm{bol}}$ could appear in the relative metallicities among the bins because more luminous quasars tend to be more metal-rich than less luminous quasars \citep{Hamann99, Warner04, Shemmer04}.  However, the absence of a corresponding gradient in the metallicity is not surprising given the relatively high average of the total L$_{\mathrm{bol}}$, the range in average L$_{\mathrm{bol}}$ among the composites of less than 0.4~dex (Table~\ref{tab:abs}), which corresponds to a range in metallicity of no more than 0.2~dex in \citet{Warner04}, and the sensitivity of the data to trends in metallicity greater than $\sim$0.3~dex between the three L$_{60}$ bins. 

The fact that the data show no significant metallicity evolution with changing SFR (L$_{60}$) could be due to the small sample sizes being affected by object-to-object scatter or by the small dynamic range in L$_{60}$, with values that are all within the luminosity/SFR range of powerful ULIRGs.  

Absorption features such as AALs and BALs and reddened continua may be more numerous among quasar environments that have undergone a recent merger or are participating in the blowout of gas that is revealing the visible quasar \citep{Becker00, Richards01a, Dodorico04, Georgakakis09}.  Although these are optically selected quasars chosen to be bright in the rest-frame UV and strongly reddened sources would be excluded, there could still be a more subtle trend in sub-samples by L$_{60}$.  We do not find a trend in the number of AALs per quasar or in the continuum slope per L$_{60}$ bin.  Each bin appears to have roughly the same percentage of quasars with AALs.  The absence of a trend could be due simply to the small numbers in our sample.  Or, if the consistency across bins is real, it  could indicate that there is no progression in quasar outflow characteristics with changing SFRs.  A larger sample size is needed to resolve this question.

The overall conclusion from this and other studies is that quasars appear to be metal-rich at all redshifts, with only small variations due to SMBH mass and luminosity dependencies \citep{Warner04, Shemmer04, Nagao06a, Jiang07, Juarez09}.  The consistently high metallicities suggest that the star formation producing these FIR-bright quasars is not the star formation that determines the metallicity of the BLR.  Instead, the BLR gas must have been enriched prior to these FIR-producing star formation episodes, possibly shortly after a merger or some other star formation trigger when the starburst first began.  There are no significant differences among the L$_{60}$ bins in the outflow fraction or in the amount of debris leftover from mergers in the form of absorption, which might correlate with L$_{60}$ if this is indeed an evolutionary sequence. This scenario is consistent with models of galaxy and quasar evolution, which predict a lack of metallicity evolution.  The models also find that quasar activity, though caused by a disruption of some sort, is significantly delayed after the disruption event and most of the star formation and enrichment in the host is complete even as the quasar phase first begins to emerge \citep{DiMatteo05, Hopkins08, Li08a}.  The FIR-bright quasars may be in some sense less mature than the FIR-faint quasars, however, this is not manifest in significant metallicity evolution over the timescales probed by this sample,  which are shorter than a quasar lifetime. 

The reality of quasar-host galaxy formation could also be more complicated than any simple monotonic sequence in which high SFRs correlate cleanly with lower metallicities and an earlier stage of evolution.  \citet{Veilleux09} suggest a `softer' ULIRG-quasar evolution paradigm with more scatter in the path of an individual galaxy's evolution from ULIRG to quasar.  They measure neon abundances of Z$\sim$2.9~Z$_\odot$ in the nuclear regions of ULIRGs.  If these high metallicities are ubiquitous in ULIRGs, we will never observe metal-poor quasars, regardless of their stage of evolution.  Other plausible scenarios suggest that AGN and star formation activity in galaxies may be episodic in nature \citep{DiMatteo08}.  Thus, as a quasar emerges at the tail end of a star-forming phase, there is no guarantee that its BLR metallicity is linked to this recent and/or ongoing star formation phase, because several previous episodes of star formation and AGN activity could already have occurred and enriched the BLR gas to supersolar values. 

An alternative interpretation is provided by the works of \citet{Dave07} and \citet{Finlator08}.  They postulate a scenario where a star-forming galaxy could quickly reach an equilibrium metallicity early in its formation history.  Higher mass galaxies would have higher abundances, but for a given galaxy, the abundance would not change significantly after an initial balance of in-flowing metal-poor and out-flowing metal-rich gas was reached.  The lack of BLR metallicity evolution with star formation rate found in this study is consistent with these objects having previously reached an equilibrium metallicity, unaffected by the observed ongoing star formation.

\section{Summary}
\label{sec:sum}

We create composite rest-frame UV spectra for a sample of 34 quasars observed in the FIR and measure \nv/\civ~ and \siiv+\oiv/\civ~ flux ratios.  We convert these flux ratios into a metallicity for each of three L$_{60}$ bins, which we interpret as SFR bins.  We find that all three bins have supersolar metallicities of the order 4 to 9 times solar, and find no metallicity trend with L$_{60}$.  We investigate the amount of outflowing gas, as manifest by both AALs and somewhat broader absorption lines, in each L$_{60}$ bin and find no difference in number between the L$_{60}$ bins, neither do we find a change in continuum slope with changing L$_{60}$, as would be consistent with more dusty star formation in hosts with higher SFRs.  Together, these results suggest that the ongoing star formation in the host is not responsible for the metal enrichment of the BLR gas.  Instead, the BLR gas must have been enriched before the visible quasar phase.  We note that we require a substantially larger sample size with a broader range in L$_{60}$ to confirm that these results are not affected by unusual individual objects or biased by the narrow L$_{60}$ range.

\bibliographystyle{mn2e}

\end{document}